\documentclass[12pt]{article}
\usepackage{graphicx}
\usepackage{dcolumn}
\usepackage{bm}
\usepackage{amsmath}
\usepackage{amssymb}
\usepackage{color}
\title{Proposed measurement of simultaneous particle and wave properties 
of electric current in a superconductor}
\author{
Hrvoje Nikoli\'c \\
Theoretical Physics Division, Rudjer Bo\v{s}kovi\'{c} Institute, \\
P.O.B. 180, HR-10002 Zagreb, Croatia \\
{\normalsize e-mail: hnikolic@irb.hr} \\
\and
Josip Atelj \\
Department of Physics, Faculty of Science, University of Zagreb, \\
10000 Zagreb, Croatia\\
{\normalsize e-mail: ateljosip@gmail.com} \\
\makebox[1in]{} \\
}
\date{\today}
\begin{document}
\maketitle
\begin{abstract}
In a microscopic quantum system 
one cannot perform a simultaneous measurement of particle and wave properties. 
This, however, may not be true for macroscopic quantum systems.
As a demonstration, we propose to measure the local macroscopic current 
passed through two slits in a superconductor. 
According to the theory based on the linearized Ginzburg-Landau equation 
for the macroscopic pseudo wave function, the streamlines of the measured 
current should have the same form as particle trajectories 
in the Bohmian interpretation of quantum mechanics.
By an explicit computation we find that the streamlines 
should show a characteristic wiggling, which is a consequence of quantum interference.
\end{abstract}


\section{Introduction} 

According to wave-particle complementarity in quantum mechanics (QM), one cannot 
simultaneously measure (with arbitrary precision) both the wave-like properties and the particle-like
properties of a microscopic quantum object. For instance, a measurement of electron's position 
necessarily ``collapses'' the wave function to a distribution well localized in space, which destroys
the wave-like properties associated with wave functions widely extended in space.
A more precise formulation of wave-particle complementarity is the Heisenberg uncertainty principle 
$\Delta x \Delta p \ge \hbar/2$ \cite{griffiths}. For instance, if the wave packet is well localized in space so that 
$\Delta x$ is small, then $\Delta p$ must be large so that the wave packet cannot be well approximated 
by a plane wave $e^{ipx/\hbar}$. A consequence is that one cannot measure the particle trajectory with arbitrary
precision, because a trajectory requires both position and velocity to be well defined.   
In spite of this, the Bohmian interpretation of QM \cite{bohm,book-bohm,book-hol,book-durr,oriols}
proposes that fundamental microscopic particles have well defined trajectories.
For all practical purposes, however, the Bohmian interpretation makes the same measurable predictions 
as standard QM and offers an explanation why, in practice, the trajectories cannot be measured directly
\cite{durrabs,nik-ibm}.

There is, however, the possibility to measure the Bohmian trajectories {\em indirectly}. 
One such possibility is to measure the trajectories with weak measurements, as proposed in \cite{wiseman}, 
further analyzed in \cite{durr-weak} and finally realized in the laboratory in \cite{weak-science}.
In this paper we propose a different possibility of indirect measurement of Bohmian trajectories,
not based on weak measurements, but based on a macroscopic quantum phenomenon - superconductivity.

The basic idea is to measure the direction of the local electric current ${\bf j}$ 
as a function of the space position ${\bf r}$. 
More specifically, we propose to study a planar conductor in the $x$-$y$ plane. 
The local electric current can be determined
experimentally by using the Hall probe nearly above the conductor to measure the magnetic field
induced by the local current. The directions of ${\bf j}$ at different positions define  
the streamlines of the electric current. As we discuss in the paper, the current can be described 
theoretically by the macroscopic Ginzburg-Landau theory \cite{gl,annett,kittel}, 
which, in the linear approximation, predicts 
that the streamlines have the same form as particle trajectories in the Bohmian interpretation of QM. 
Some relations between Ginzburg-Landau theory and the Bohmian interpretation have also been
discussed in \cite{berger,nikulov}.

We stress that such indirect measurements are not aimed to prove that 
the Bohmian interpretation of QM is right. The Bohmian interpretation
claims that each individual particle follows such a trajectory, while indirect measurements 
involve some sort of averaging over many particles, revealing no direct information on the behavior 
of individual particles. 
In the case of weak measurements one averages 
over many repetitions of a measurement at a given ${\bf r}$, each time with another particle 
prepared in the same way. 
In the case of superconductivity a local measurement is really a measurement 
in a region of a macroscopic size (typically of the order of 1 mm) containing many microscopic particles. 
Indeed, all such measurements can also be explained with the standard ``Copenhagen'' interpretation of QM. 
Nevertheless, such measurements can demonstrate that there is something measurable that follows
a trajectory that looks exactly like a Bohmian trajectory.


\section{Elements of Ginzburg-Landau theory}

The Ginzburg-Landau theory is a phenomenological macroscopic theory of superconductivity in which the basic
entity is the macroscopic complex valued field $\psi({\bf r})$ \cite{gl,annett,kittel}. 
This field is the order parameter associated with the phase transition at the critical 
temperature $T_c$ at which the system becomes superconductive. For temperatures $T>T_c$ 
the ordering parameter vanishes $\psi=0$, while for $T<T_c$ it satisfies the Ginzburg-Landau equation
\begin{equation}\label{gl}
 -\frac{\hbar^2}{2m} \left(\! \mbox{\boldmath $\nabla$}-\frac{iq}{\hbar c}{\bf A}({\bf r}) \!\right)^2 \!\psi({\bf r})
+(-a+b|\psi({\bf r})|^2) \psi({\bf r}) =0. 
\end{equation}
Here $m=2m_e$ and $q=-2e$ are the effective mass and effective charge, respectively, of the Cooper pair,
${\bf A}({\bf r})$ is the vector potential of the external magnetic field 
${\bf B}({\bf r})=\mbox{\boldmath $\nabla$}\times {\bf A}({\bf r})$, 
$c$ is the speed of light, while
$a$ and $b$ are phenomenological parameters
that depend on temperature. Close to $T_c$ the two parameters have the expansions 
$a(T)=-a_1(T-T_c)+\ldots$, $b(T)=b_0+\ldots$, 
where $a_1$ and $b_0$ are positive constants so that $a$ and $b$ are positive in the superconducting phase.     
The main physical quantity derived from $\psi({\bf r})$ is the local current ${\bf j}({\bf r})$
given by 
\begin{equation}\label{jgl}
 {\bf j}=-i\frac{q\hbar}{2m} [\psi^*(\mbox{\boldmath $\nabla$}\psi)-(\mbox{\boldmath $\nabla$}\psi^*)\psi]
-\frac{q^2}{mc}\psi^*\psi {\bf A} .
\end{equation}
Eq.~(\ref{gl}) implies that (\ref{jgl})
obeys local conservation $\mbox{\boldmath $\nabla$}{\bf j}=0$.
The current (\ref{gl}) is interpreted as the local electric current density in the superconductor,
while $\psi^*\psi\equiv n$ is interpreted as the concentration of quasiparticles (Cooper pairs)
that carry the current. 
Hence the local velocity of the current carriers is
\begin{equation}\label{v}
 {\bf v}({\bf r})=\frac{{\bf j}({\bf r})}{q n({\bf r})} .
\end{equation}
Writing $\psi$ in the polar form $\psi({\bf r})=|\psi({\bf r})|e^{i\varphi({\bf r})}$, (\ref{v})
can be written as
\begin{equation}\label{v2}
 {\bf v}({\bf r})=\frac{1}{m}\left( \hbar\mbox{\boldmath $\nabla$}\varphi({\bf r})-\frac{q}{c}{\bf A}({\bf r}) \right) .
\end{equation}

In some cases, the non-linear term $b|\psi|^2$ in (\ref{gl}) can be neglected.
In this limit, (\ref{gl}) takes the same form as 
a time-independent Schr\"odinger equation. Nevertheless, the interpretation is different.
In QM of a single particle, $|\psi({\bf r})|^2$ is interpreted as the probability density of particle 
to be found at the position ${\bf r}$. This probabilistic interpretation is closely related
to wave function ``collapse'', which can be thought of as an update of knowledge about the 
particle position achieved by measurement. This ``collapse'', that is a change of wave function by the process
of measurement, lies at the heart of wave-particle complementarity and the Heisenberg uncertainty principle.   
In (\ref{gl}), however, $|\psi({\bf r})|^2$ is not interpreted as a probability, so measurement
is not associated with a wave function collapse. Instead, it is interpreted as a macroscopic density
originating form a large number of Cooper pairs. To emphasize that $\psi$ in (\ref{gl})
has a different physical interpretation than $\psi$ in the single-particle Schr\"odinger equation,
$\psi$ in (\ref{gl}) is often referred to as pseudo wave function. 
With the hope that it will not raise any confusion, in the rest of the paper we shall refer to $\psi$ in (\ref{gl})
simply as wave function. 

\section{Relation with the Bohmian interpretation}

Eq.~(\ref{v2}) has the same form as the formula for particle velocity in the Bohmian interpretation of QM 
\cite{book-bohm,book-hol}. Moreover, neglecting the non-linear term and following Feynman \cite{feynman}
one finds that the acceleration $d{\bf v}/dt$ satisfies
\begin{equation}\label{feynman}
 m\frac{d{\bf v}}{dt}=\frac{q}{c}{\bf v}\times{\bf B}- \mbox{\boldmath $\nabla$} Q ,
\end{equation}
where
\begin{equation}\label{Q}
 Q=-\frac{\hbar^2}{2m}\frac{\mbox{\boldmath $\nabla$}^2 |\psi|}{|\psi|} .
\end{equation}
The first term on the right-hand side of (\ref{feynman}) is the classical magnetic force, 
while the second term is a quantum force determined by the quantum potential (\ref{Q}).
The formula (\ref{Q}) has the same form as the formula for the quantum potential
in the Bohmian interpretation of QM \cite{bohm,book-bohm,book-hol}. 
The same formula for the quantum potential appears also in the old Madelung hydrodynamic
interpretation \cite{madelung} of the Schr\"odinger equation.

\section {Modeling the wave function}

Our goal now is to find $\psi({\bf r})$ that models a configuration in a realistic experiment.
We study a superconductor in the absence of an external magnetic field, so ${\bf A}=0$
and (\ref{gl}) simplifies to
\begin{equation}\label{glA=0}
 \mbox{\boldmath $\nabla$}^2\psi({\bf r})+\kappa^2(1-\beta|\psi({\bf r})|^2)\psi ({\bf r})= 0 ,
\end{equation}
where
\begin{equation}
\kappa^2=\frac{2ma}{\hbar^2}, \;\;\; \beta=\frac{b}{a} .
\end{equation}
Here ${\bf r}={\hat{\bf x}}x+{\hat{\bf y}}y+{\hat{\bf z}}z$, where $(x,y,z)$ are Cartesian coordinates
and $({\hat{\bf x}},{\hat{\bf y}}, {\hat{\bf z}})$ are the corresponding unit vectors.
We consider a planar superconductor in the $x$-$y$ plane, so in the rest of the analysis nothing depends on $z$. 

As a first step, we find the solution of (\ref{glA=0}) 
that describes a current in the $x$-direction traveling from the left to the right.
Searching for a plane-wave solution $\psi(x)=\sqrt{n_0}e^{ikx}$, where $n_0$ is a constant 
concentration while $k$ is real and positive, we find that (\ref{glA=0}) is satisfied provided that
\begin{equation}\label{k2}
 k^2=\kappa^2(1-\beta n_0) .
\end{equation}
Hence the maximal value of $k$ is $k_{\rm max}=\kappa$. From (\ref{v2}) we see that the velocity 
in the $x$-direction is $v=\hbar k/m$, so the maximal velocity is $v_{\rm max}=\hbar\kappa/m$. 
Hence (\ref{k2}) implies that the non-linear term $\beta|\psi|^2$ in (\ref{glA=0}) can be written as
\begin{equation}
 \beta n_0 = \frac{\kappa^2-k^2}{\kappa^2}= \frac{v_{\rm max}^2-v^2}{v_{\rm max}^2},
\end{equation}
implying that the non-linear term can be neglected when $k$ is close to its maximal value $\kappa$.  

\begin{figure}[t]
\includegraphics[width=8cm]{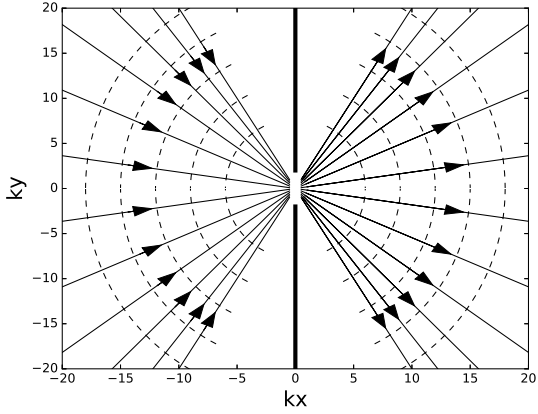}
\caption{\label{fig1}
The current streamlines (full lines with arrows) in a single slit experiment. The dashed curves 
are the corresponding wave fronts perpendicular to the streamlines.}
\end{figure}

In the next step we insert a barrier along the $y$-axis at $x=0$, with a single slit 
in the barrier drilled at $x=y=0$, as in Fig.~\ref{fig1}.
On the right from the slit, that is for $x>0$, the slit effectively looks like a ``source''
of the wave. But the wave for $x>0$ really originates from the wave hitting the slit
from the left, which means that for $x<0$ the slit effectively looks like a ``sink''.
We are not interested in the region near the barrier, so we can use an approximation 
in which the wave fronts spread in concentric semicircles for $x>0$ and 
shrink in concentric semicircles for $x<0$. In this approximation 
$\psi$ depends only on $r\equiv\sqrt{x^2+y^2}$, 
so we use cylindrical coordinates $(r,\varphi,z)$
and write $\psi({\bf r})=\psi(r)$, implying that (\ref{glA=0}) reduces to 
\begin{equation}\label{helmr}
  \frac{\partial^2\psi}{\partial r^2} +\frac{1}{r}\frac{\partial\psi}{\partial r} 
+\kappa^2(1-\beta|\psi|^2)\psi =0 .
\end{equation}
With the ansatz 
\begin{equation}\label{anz}
 \psi(r)=f(r)\frac{e^{ikr}}{\sqrt{r}} ,
\end{equation}
(\ref{helmr}) reduces to the non-linear equation for $f$
\begin{equation}\label{eqf}
 f''+2if'+\left( \frac{1}{4\tilde{r}^2} +\frac{\kappa^2-k^2}{k^2} 
-\frac{\kappa^2}{k}\frac{\beta|f|^2}{\tilde{r}} \right)f =0,
\end{equation}
where $\tilde{r}=kr$ is the dimensionless radial coordinate 
and the primes denote derivatives over $\tilde{r}$. 
This equation can be solved numerically, but we find it more illuminating
to give an approximative analytic solution. Trying the ansatz $f(r)=f_0={\rm constant}$,
we see that (\ref{eqf}) is approximately satisfied if the bracket in (\ref{eqf}) is small.
Hence $f(r)=f_0$ is a good approximation when (i) $k$ is close to its 
maximal value $\kappa$ and (ii) $\tilde{r}\gg 1$. In this limit, in particular, 
the non-linear term $\beta|f|^2/\tilde{r}$ is negligible.
The value of $f_0$ will not matter for computation of 
the streamlines, so for convenience we take $f_0=1$. In this way 
we see that an approximative solution of (\ref{helmr}) is 
\begin{equation}\label{psir}
 \psi(r)=\frac{e^{ikr}}{\sqrt{r}} .
\end{equation}
Since $k$ is defined as positive, (\ref{psir}) describes a radial outgoing stream 
for $x>0$. For $x<0$ we have a radial ingoing stream $e^{-ikr}/\sqrt{r}$. Hence the full solution, 
within our approximations, is
\begin{equation}\label{psipm}
\psi(r)=\left\{
\begin{array}{c}
\displaystyle\frac{e^{ikr}}{\sqrt{r}} \;\; {\rm for} \;\; x>0 \\
\displaystyle\frac{e^{-ikr}}{\sqrt{r}} \;\; {\rm for} \;\; x<0 .
\end{array}
\right.
\end{equation}

Now consider two slits in the barrier. The barrier is again positioned along the $y$-axis at $x=0$. 
We put slit-1 at $y=d/2$ and slit-2 at $y=-d/2$, where $d$ is the distance between the slits.
When only slit-1 [or only slit-2] is open, then the wave function is $\psi_1(x,y)=\psi(r_1)$
[or $\psi_2(x,y)=\psi(r_2)$], where $\psi(r)$ is given by (\ref{psipm}) and
\begin{eqnarray}
& r_1(x,y)=\sqrt{x^2+(y-d/2)^2} , & 
\nonumber \\
& r_2(x,y)=\sqrt{x^2+(y+d/2)^2} . &
\end{eqnarray}
To see what happens when both slits are open, we recall that (\ref{psipm}) 
has been obtained in the regime in which the non-linear term can be neglected.
Hence, in this regime, we can use the superposition principle,
so the wave function when both slits are open can be taken to be
\begin{equation}\label{Psi}
 \Psi(x,y) = \frac{1}{\sqrt{2}}(\psi(r_1)+\psi(r_2)) ,
\end{equation}
where $\psi(r)$ is given by (\ref{psipm}).

\section{Computation of streamlines} 

\begin{figure*}[t]
\includegraphics[width=12cm]{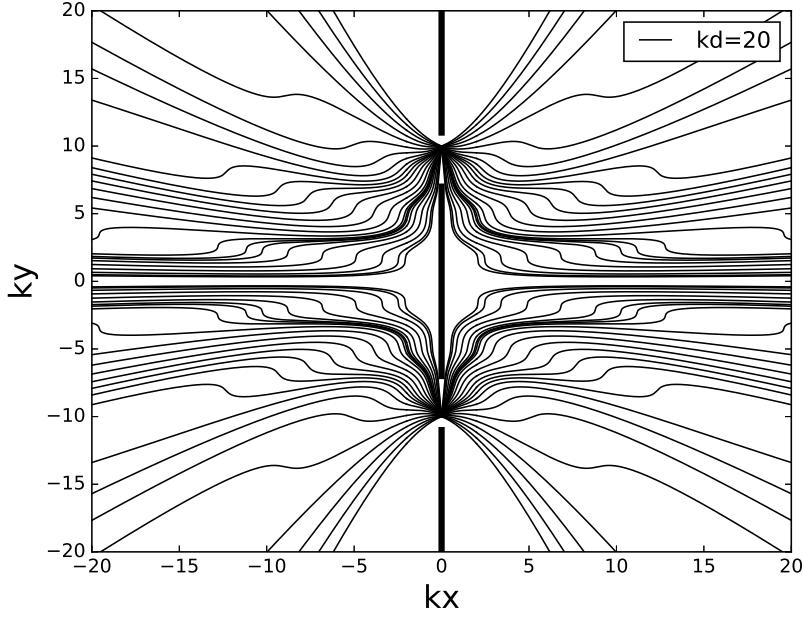}
\caption{\label{fig2}
The streamlines of current traveling through two slits for $kd=20$. 
}
\end{figure*}

Now the current is given by the formula (\ref{jgl})
with $\psi\rightarrow\Psi$ and ${\bf A}=0$. Since the constant $q\hbar/2m$ in (\ref{jgl}) does not matter 
for computation of the streamlines, we take
\begin{equation}\label{j}
 {\bf j}=-i[\Psi^*(\mbox{\boldmath $\nabla$}\Psi)-(\mbox{\boldmath $\nabla$}\Psi^*)\Psi] .
\end{equation}
Inserting (\ref{Psi}) into (\ref{j}), after a straightforward calculus we obtain
\begin{equation}\label{jfin}
 {\bf j}=\frac{{\rm sign}(x) k}{\sqrt{r_1r_2}} 
(g_1\mbox{\boldmath $\nabla$}r_1+g_2\mbox{\boldmath $\nabla$}r_2) ,
\end{equation}
where 
\begin{eqnarray}\label{jfinaux}
 & g_1=\displaystyle\sqrt{\frac{r_2}{r_1}} + \cos\varphi_{21} + \frac{\sin\varphi_{21}}{2kr_1} ,&
\nonumber \\
& g_2=\displaystyle\sqrt{\frac{r_1}{r_2}} + \cos\varphi_{21} - \frac{\sin\varphi_{21}}{2kr_2} , &
\\
& \varphi_{21}=k(r_2-r_1) , &
\\
\label{jfinaux2}
& \mbox{\boldmath $\nabla$}r_1= \hat{\bf x} \displaystyle\frac{x}{r_1} + \hat{\bf y} \frac{y-d/2}{r_1} , &
\nonumber \\
& \mbox{\boldmath $\nabla$}r_2= \hat{\bf x} \displaystyle\frac{x}{r_2} + \hat{\bf y} \frac{y+d/2}{r_2} , &
\end{eqnarray}
and ${\rm sign}(x)=\pm 1$ for $x\gtrless 0$. 
The streamlines in the $x$-$y$ plane can be computed by numerical integration of
$dy/dx=j_y/j_x$, or equivalently
\begin{equation}\label{streamfin}
 \frac{dy}{dx}=\frac{m_y}{m_x} ,
\end{equation}
where ${\bf m}\equiv g_1\mbox{\boldmath $\nabla$}r_1+g_2\mbox{\boldmath $\nabla$}r_2$.
In numerical integration we use dimensionless coordinates $\tilde{x}=kx$, $\tilde{y}=ky$.
The results are shown in Fig.~\ref{fig2}. The streamlines show characteristic wiggling 
typical for Bohmian trajectories in two slit configurations \cite{philippidis79,philippidis82,book-hol}.
This wiggling is a consequence of quantum interference, or equivalently, of the quantum force
described by (\ref{feynman}). In this way Fig.~\ref{fig2} shows a simultaneous particle-like and
wave-like macroscopic properties of the electric current in the superconductor.

\section{Conclusion} 

In this paper we have modeled the Landau-Ginzburg wave function 
describing the electric current in a superconductor passed through two slits.
From the wave function we have computed the streamlines of the current.
Those streamlines can also be determined experimentally, by using the Hall probe to 
measure the local direction of the current. Our computation is based on a linear approximation,
which is expected to be a good approximation far from the slits where the quantum interference 
effects are pronounced. As a consequence of quantum interference, our computation shows 
a characteristic wiggling of the streamlines, typical for quantum trajectories 
in the Bohmian interpretation of quantum mechanics. Experimental confirmation of such wiggling 
would be a demonstration that the macroscopic electric current in a superconductor 
can show both particle properties and wave properties simultaneously.  

\section*{Acknowledgments}

H.N. is grateful to D. \v{C}apeta, Z. Ere\v{s} and V. Zlati\'c for discussions. 
The work of H.N. was supported  
by the European Union through the European Regional Development Fund 
- the Competitiveness and Cohesion Operational Programme (KK.01.1.1.06).

\end{document}